\documentstyle[aps,prl,epsfig, twocolumn]{revtex}

\begin{document}
\title{Quantum entanglement in spinor Bose-Einstein condensates}
\author{L.-M. Duan$^{1,2}$, J. I. Cirac$^{1}$, and P. Zoller$^{1}$}
\address{$^{1}$Institut f\"{u}r Theoretische Physik, Universit\"{a}t Innsbruck,
A-6020 Innsbruck, Austria \\
$^{2}$Laboratory of Quantum Communication and Computation, USTC, Hefei
230026, China}
\maketitle

\begin{abstract}
We propose a scheme to generate and detect various kinds of quantum
entanglement in a spin-1 Bose-Einstein condensate. It is shown that
substantial many-particle entanglement can be generated directly in the
spin-1 condensate by free dynamical evolution with a properly prepared
initial state. The scheme also provides a simple method to generate
three-mode entanglement in the second-quantization picture and to detect the
continuous variable type of entanglement between two effective modes in the
spin-1 condensate.
\end{abstract}

The achievement of atomic Bose-Einstein condensate (BEC) [1] has raised
extensive experimental and theoretical studies in this area, and these
studies have opened up some possibilities for important applications. An
impressive example of these applications is to use BECs with internal
degrees of freedom for generation of quantum entanglement [2-7], which is
the essential ingredient of many quantum information protocols [8]. To
generate entanglement, one requires an experimental system which can be
prepared in a pure state, and there should be significant interactions
between the relevant particles which can be coherently controlled. BECs
fulfil these requirements, and therefore they could provide an ideal
experimental system for studying quantum entanglement. It has been shown
that with coherent collision interactions in BECs, it is possible to
generate substantial many-particle entanglement in a two-component
condensate [2], or to generate entangled atomic beams from the outputs of a
single condensate prepared in a specific internal level which enables
spin-exchange collisions [3,4].

A good experimental system for demonstration of these proposals is the
Sodium condensate confined in an optical trap [9-11]. Recently, Stamper-Kurn 
{\it et al.} have realized an optically trapped BEC in which all the three
ground Zeeman levels of sodium atoms are involved [9]. In such a
three-component (so-called spin-1) condensate, there are spin-exchange
collisions, which constantly mix the different spin components, and other
kinds of internal spin dynamics, which result in complex nonlinear collision
phase shifts [12-17]. In this paper, we will investigate whether one can
directly prepare and detect quantum entanglement in such a spin-1
condensate. The motivation of this work is two-folds: firstly, direct
detection of quantum entanglement in the realized spin-1 condensate
eliminates the requirements of the level shifting and the output techniques
in previous works [2-4], and thus allows for a easier experiment; and
secondly, the spin-1 condensate with the presence of both the spin-exchange
collisions and the nonlinear phase shifts is a more complicated system, and
opens up possibilities for richer physical behaviors. We will show that
various kinds of entanglements are available in this system. We can get
substantial many-particle entanglement (the spin squeezing type of
entanglement [18,19]) as well as three-mode and two-mode entanglement (the
continuous-variable type of entanglement [20]). All these entanglements are
useful for different purposes of applications.

The spin-1 BEC has been studied both theoretically [12-17] and
experimentally [9-11], with the emphasis on the ground-state structure, the
spin-mixing dynamics, and the response to the external magnetic field. Here,
we will focus on its entanglement property. In Ref. [15], it was shown as a
by-product that the spin ground state of the sodium condensate with the
anti-ferromagnetic interaction should be entangled. Unfortunately, this
ground-state entanglement was not confirmed by the experiment [10] because
it is very hard to cool the system to the spin ground state (the energy gap
between the spin ground and the first excited states is many orders smaller
than the chemical potential), and the spin ground state is very fragile
under the influence of the external magnetic field [17]. Here, we describe a
different method to generate and detect quantum entanglement by free
dynamical evolution with a proper initial state. As we will see, this
provides a feasible method of detecting entanglement in the spin-1 BEC with
the current technology.

Before we enter the discussion of the spin-1 condensate, let us first
clarify different kinds of entanglements available in a bosonic
many-particle system. Consider $N$ bosonic identical particles with each
particle having three internal levels $\left| +1\right\rangle $, $\left|
-1\right\rangle $, and $\left| 0\right\rangle $. First, the $N$ particles
are said to be entangled in the first quantization picture if their internal
density matrix $\rho $ cannot be decomposed into 
\begin{equation}
\rho =\sum_{k}p_{k}\rho _{1}^{k}\otimes \rho _{2}^{k}\otimes \cdots \otimes
\rho _{N}^{k}\text{,}
\end{equation}
where the coefficients $p_{k}$ are positive real numbers satisfying $%
\sum_{k}p_{k}=1$, and $\rho _{i}^{k}$ is a $3\times 3$ density matrix of the 
$i$th particle. To check whether an $N$-particle system is entangled, a good
experimentally detectable criterion is the squeezing parameter. For
particles with only two internal levels $\left| a\right\rangle $ and $\left|
b\right\rangle $, one can construct a collective spin operator ${\bf J}%
_{ab}=\sum_{i=1}^{N}{\bf j}_{i}$, with the individual spin-$1/2$ operators $%
j_{i}^{x}=\left( \left| a\right\rangle _{i}\left\langle b\right| +\left|
b\right\rangle _{i}\left\langle a\right| \right) /2,$ $j_{i}^{y}=i\left(
\left| a\right\rangle _{i}\left\langle b\right| -\left| b\right\rangle
_{i}\left\langle a\right| \right) /2,$ and $j_{i}^{z}=\left( \left|
b\right\rangle _{i}\left\langle b\right| -\left| a\right\rangle
_{i}\left\langle a\right| \right) /2$. The squeezing parameter $\xi _{ab}^{%
{\bf n}}$ is defined directly from the variance and mean values of the ${\bf %
J}_{ab}$ operator 
\begin{equation}
\xi _{ab}^{{\bf n}}=\frac{N\left\langle \left( \Delta J_{ab}^{n_{1}}\right)
^{2}\right\rangle }{\left\langle J_{ab}^{n_{2}}\right\rangle
^{2}+\left\langle J_{ab}^{n_{3}}\right\rangle ^{2}},
\end{equation}
where $J_{ab}^{n}={\bf n}\cdot {\bf J}_{ab}$, and the ${\bf n}$s are
mutually orthogonal unit vectors. For two-level particles, the spin
squeezing parameter (2) with a value less than $1$ provides a sufficient
condition for many-particle entanglement [2] (such a case is also referred
as spin squeezing [18,19]). For three-level particles, we can construct two
orthonormal states $\left| a\right\rangle $ and $\left| b\right\rangle $
from arbitrary superpositions of the three levels $\left| +1\right\rangle $, 
$\left| -1\right\rangle $, and $\left| 0\right\rangle $, and define the $%
{\bf J}_{ab}$ and $\xi _{ab}^{{\bf n}}$ in the same way. Then, $\xi _{ab}^{%
{\bf n}}<1$ assures that the projected $N$-particle density matrix $P\rho P$
onto the subspace spanned by the states $\left| a\right\rangle _{i}$ and $%
\left| b\right\rangle _{i}$ ($i=1,2,\cdots ,N$) is entangled. Since the
projection $P$ is a local operator which cannot increase entanglement, the
original density matrix $\rho $ for the $N$ three-level particles should
also be entangled if $\xi _{ab}^{{\bf n}}<1$. For our later applications, we
define the spin operators ${\bf J}_{\pm }$ and the squeezing parameters $\xi
_{\pm }^{{\bf \theta }}$ respectively by choosing $\left| a\right\rangle
=\left| \pm \right\rangle =\left( \left| +1\right\rangle \pm \left|
-1\right\rangle \right) /\sqrt{2}$, $\left| b\right\rangle =\left|
0\right\rangle $, and $n_{1}=\left[ \cos \theta ,\sin \theta ,0\right] $ in
Eq. (2). Either of $\xi _{\pm }^{{\bf \theta }}$ less than $1$ for some $%
\theta $ is a signature of many particle entanglement. Note that the spin
squeezing type of entanglement represented by a small $\xi _{ab}^{{\bf n}}$
is useful in practice since the inverse of $\xi _{ab}^{{\bf n}}$ gives
exactly the improvement of the single-to-noise ratio for atomic clocks
relative to the standard quantum limit with the Ramsey scheme involving the
levels $\left| a\right\rangle $ and $\left| b\right\rangle $ [19].

The above system can also be described in the second quantization picture.
We have three bosonic modes $a_{\alpha }$ $\left( \alpha =0,\pm 1\right) $
corresponding to the three internal levels, with the commutation relations $%
\left[ a_{\alpha },a_{\beta }^{\dagger }\right] =\delta _{\alpha \beta }$.
These modes can also be entangled like the individual particles. Similar to
the definition in Eq. (1), we say that two modes $a_{\alpha },a_{\beta }$
are entangled if their density matrix cannot be expressed as $%
\sum_{k}p_{k}\rho _{\alpha }^{k}\otimes \rho _{\beta }^{k}$, and all the
three modes $a_{\alpha },a_{\beta },a_{\gamma }$ are intrinsically entangled
if their density matrix cannot be expressed as $\sum_{k}p_{k}\rho _{\alpha
}^{k}\otimes \rho _{\beta \gamma }^{k}$, where $\alpha ,\beta ,\gamma =0,\pm
1$, $p_{k}$ are positive real numbers, and $\rho _{\alpha \left( \beta
,\beta \gamma \right) }^{k}$ is the a density matrix of the corresponding
mode(s). Note that the mode entanglement in the second quantization and the
particle entanglement in the first quantization are two different concepts,
and they are useful for different purposes of applications. One can easily
find example states which are mode entangled but not particle entangled, or
particle entangled but not mode entangled [22]. Mode entanglement of bosonic
atoms could be useful in various continuous variable or qubit
quantum information processing schemes currently based on entangled optical
modes [8]. For two modes $a_{\pm 1}$, there is an easy experimentally
detectable entanglement criterion [20]. One can define quadrature phase
amplitudes $X_{\pm 1}^{\theta }=\left( e^{i\theta }a_{\pm 1}^{\dagger
}+e^{-i\theta }a_{\pm 1}\right) /\sqrt{2}$ and $X_{\pm }^{\theta }=\left(
X_{+1}^{\theta }\pm X_{-1}^{\theta }\right) /\sqrt{2}$ for the two modes $%
\pm 1$. It has been shown in Ref. [20] that $\left\langle \left( \Delta
X_{+}^{\theta }\right) ^{2}\right\rangle +\left\langle \left( \Delta
X_{-}^{\theta +\pi /2}\right) ^{2}\right\rangle <1$ for some angle $\theta $
provides a criterion for entanglement between the modes $a_{+1}$ and $a_{-1}$
(a state with this property is also referred as a two-mode squeezed state,
or a continuous variable entangled state [8,20]). For later applications in
the spin-$1$ condensate, we are more interested in two effective bosonic
modes $a_{\pm 1}^{\prime }$, defined as $a_{\pm 1}^{\prime }=a_{\pm
1}a_{0}^{\dagger }/\sqrt{\left\langle a_{0}^{\dagger }a_{0}\right\rangle }$.
The modes $a_{\pm 1}^{\prime }$ are approximately bosonic if nearly all the
atomic populations remain in the component $a_{0}$, i.e., $\left\langle
a_{0}^{\dagger }a_{0}\right\rangle \simeq N$. The definition of the new
modes $a_{\pm 1}^{\prime }$, which are proportional to $a_{\pm 1}$, is to
remove the phase randomness in the condensate by phase locking both of the
modes $a_{\pm 1}$ to the condensate mode $a_{0}$. For the modes $a_{\pm
1}^{\prime }$, the quadrature phase amplitudes defined above reduce to $%
\left\langle \left( \Delta X_{\pm }^{\theta }\right) ^{2}\right\rangle
\simeq \xi _{\pm }^{{\bf \theta }}/2$. The inequality $\xi _{+}^{{\bf \theta 
}}+\xi _{-}^{{\bf \theta }}<2$ from the spin squeezing parameters thus
becomes a sufficient criterion for entanglement between the modes $a_{\pm
1}^{\prime }$. As a result, a measurement of the spin squeezing parameters $%
\xi _{\pm }^{{\bf \theta }}$ provides criteria for both many-particle
entanglement and two-mode entanglement.

We will see that three-mode entanglement is also produced in a spin-1
condensate. A general criterion for the three-mode entanglement is still
absent, but our system to a good approximation will be in a pure three-mode
state $\left| \phi \right\rangle _{+1,-1,0}$, and in this case the
three-mode entanglement can be easily characterized by the quantity $E_{3}=%
\sqrt[3]{E\left( \rho _{+1}\right) E\left( \rho _{-1}\right) E\left( \rho
_{0}\right) }$, where $E\left( \rho _{\alpha }\right) =-tr\left( \rho
_{\alpha }\log _{2}\rho _{\alpha }\right) $ denotes the von Neuman entropy
of $\rho _{\alpha }$, and $\rho _{\alpha }$ is the reduced density operator
of the mode $\alpha $ $\left( \alpha =+1,-1,0\right) $. It is obvious that
for pure states $E_{3}$ is non-increasing under local operations and
positive if and only if the system state is intrinsically three-mode
entangled. It will be shown that in a spin-1 condensate $E_{3}$ can be
significantly larger than $0$. Note that different to the spin squeezing
parameters it is not easy to directly measure $E_{3}$.

Now we turn to the physical system, which is a dilute gas of trapped bosonic
atoms with hyperfine spin $f=1$. Assume that the system is very cold so that
the collision interaction between atoms is effectively described by a
pair-wise pseudo $\delta $-potential, which preserves the hyperfine spin of
individual atoms and is rotationally invariant in the hyperfine spin space
[12]. With this symmetry assumption, the most general form of the
Hamiltonian can be dived into two parts $H=H_{s}+H_{a}$, with the symmetric
and asymmetric Hamiltonians $H_{s}$ and $H_{a}$ respectively given by
[12,15] (setting $\hbar =1$) 
\begin{eqnarray}
H_{s} &=&\sum_{\alpha }\int d{\bf r}\Psi _{\alpha }^{\dagger }\left( -\frac{%
\bigtriangledown ^{2}}{2m}+V\right) \Psi _{\alpha }  \nonumber \\
&&+\frac{\lambda _{s}}{2}\sum_{\alpha ,\beta }\int d{\bf r}\Psi _{\alpha
}^{\dagger }\Psi _{\beta }^{\dagger }\Psi _{\alpha }\Psi _{\beta },
\end{eqnarray}
\begin{equation}
H_{a}=\frac{\lambda _{a}}{2}\sum_{\alpha ,\beta }\int d{\bf r}\Psi _{\alpha
}^{\dagger }\Psi _{\alpha ^{\prime }}^{\dagger }{\bf S}_{\alpha \beta }\cdot 
{\bf S}_{\alpha ^{\prime }\beta ^{\prime }}\Psi _{\beta }\Psi _{\beta
^{\prime }},
\end{equation}
where $\Psi _{\alpha }$ $\left( \alpha =-1,0,+1\right) $ is the atomic field
operator associated with atoms in the spin state $\left| f=1,m_{f}=\alpha
\right\rangle $. The first and the second terms of the Hamiltonian $H_{s}$
denote the kinetic energy and the potential energy, with $m$, the mass of
the atom, and $V$, the trap potential, which has been assumed to be the same
for all the three components. The $3\times 3$ spin matrices ${\bf S}$ in $%
H_{a}$ denote the conventional $3$-dimensional representation (corresponding
to the spin $f=1$) of the angular momentum operator, with $S_{\alpha \beta
}^{x}=\left( \delta _{\alpha ,\beta -1}+\delta _{\alpha ,\beta +1}\right) /%
\sqrt{2}$, $S_{\alpha \beta }^{y}=i\left( \delta _{\alpha ,\beta -1}-\delta
_{\alpha ,\beta +1}\right) /\sqrt{2}$, $S_{\alpha \beta }^{z}=\alpha \delta
_{\alpha \beta }$. The interaction energy has been divided into the
symmetric and the asymmetric parts, represented respectively by the last
term of $H_{s}$ and the Hamiltonian $H_{a}$. The symmetric interaction
remains the same if we exchange two arbitrary spin components, whereas the
asymmetric one is spin-dependent. The symmetric and the asymmetric collision
coefficients $\lambda _{s}$ and $\lambda _{a}$ are determined respectively
by the average and the difference of the scattering lengths in two different
collision channels. For sodium atoms, $\lambda _{a}$ is positive and about $%
25$ times smaller than $\lambda _{s}$ [11,12]. Thus, the Hamiltonian $H_{s}$
is the dominant one for determining the dynamics of the motional state.
However, the internal spin dynamics which we are interested in is completely
determined by $H_{a}$\ since the symmetric part \ $H_{s}$ is
spin-independent.

For generating entanglement, we start with a BEC\ prepared in the internal
level $\left| m_{f}=0\right\rangle $. To quantify the amount of various
kinds of squeezing and entanglement obtainable in this system, we need to
calculate the time evolution of the entanglement criteria $\xi _{\pm }^{{\bf %
\theta }}$ and $E_{3}$ defined before. For this purpose, we make use of the
single motional mode approximation to the Hamiltonian $H_{a}$ following
Refs. [15-17]. Note that the Hamiltonian $H_{s}$ is the dominant one for
determining the motional state. When the system is at a very low
temperature, the motional state is frozen to be approximately the ground
state $\varphi \left( {\bf r}\right) $ (normalized as $\int d{\bf r}\left|
\varphi \left( {\bf r}\right) \right| ^{2}=1$) of the spin-independent
Hamiltonian $H_{s}$, and we can factorize the field operators as $\Psi
_{\alpha }\approx a_{\alpha }\varphi \left( {\bf r}\right) $ $\left( \alpha
=0,\pm 1\right) $, where $a_{\alpha }$ is the usual bosonic annihilation
operator. With this approximation, one can introduce a spin operator ${\bf L}
$ with ${\bf L=}\sum_{\alpha ,\beta }a_{\alpha }^{\dagger }{\bf S}_{\alpha
\beta }a_{\beta }$, and the spin-dependent Hamiltonian $H_{a}$ is simplified
to $H_{a}=\lambda _{a}^{\prime }\left( {\bf L}^{2}-2N\right) $, where $%
\lambda _{a}^{\prime }=\lambda _{a}\int d{\bf r}\left| \varphi \left( {\bf r}%
\right) \right| ^{4}/2$. This Hamiltonian has been used for studying the
spin ground state structure [15]. We use it here for calculating squeezing
and entanglement in this system. For this purpose, it is convenient to
express the Hamiltonian $H_{a}$ by the more relevant spin operators ${\bf J}%
_{\pm }$ defined before. For the specific system configuration considered
here, $L^{z}=0$ for the initial state, and it will remain zero since $L^{z}$
is conserved. On the other hand, by definition we have $L^{x}=2J_{+}^{x}$
and $L^{y}=2J_{-}^{y}$. Therefore, the Hamiltonian $H_{a}$ expressed by $%
{\bf J}_{\pm }$ has the form 
\begin{equation}
H_{a}=4\lambda _{a}^{\prime }\left( J_{+}^{x2}+J_{-}^{y2}\right) ,
\end{equation}
where we have dropped the irrelevant constant $2\lambda _{a}^{\prime }N$.
The only non-zero commutation relation between ${\bf J}_{+}$ and ${\bf J}%
_{-} $ is $\left[ J_{+}^{x},J_{-}^{x}\right] =\left[ J_{+}^{y},J_{-}^{y}%
\right] =\left( a_{-1}^{\dagger }a_{+1}-a_{+1}^{\dagger }a_{-1}\right) /4$,
which remains very small if the population of the atoms is still dominantly
in the level $\left| m_{f}=0\right\rangle $, as it is in our scheme. In this
case, as an approximation we can consider separately the dynamics of ${\bf J}%
_{+}$ and ${\bf J}_{-}$, respectively with the effective Hamiltonian $%
J_{+}^{x2}$ or $J_{-}^{y2}$. It is well known that the Hamiltonian $%
J_{+}^{x2}$ or $J_{-}^{y2}$ will produce a spin squeezing state, with the
minimum squeezing parameter about $\left( 3/N\right) ^{2/3}/2$ after an
evolution time proportional to $N^{-2/3}/\lambda _{a}^{\prime }$ [18]. This
is the intuitive reason why we can get substantial spin squeezing and
entanglement in this system. To be more quantitative, we have calculated
exactly the time evolution of the squeezing parameter $\xi _{\pm }^{\theta }$
under the Hamiltonian (5) using the numerical approach. The system spin
state $\left| \phi \right\rangle $ under the Hamiltonian (5) can be expanded
as $\left| \phi \right\rangle =\sum_{n}c_{n}\left| N-2n,n,n\right\rangle
_{0,+1,-1}$ in the number basis, where the subscript $\alpha $ $\left(
\alpha =0,\pm 1\right) $ represents the corresponding internal mode $%
a_{\alpha }$. One can solve numerically the time evolution of the expansion
coefficients $c_{n}$, from which the spin squeezing parameters $\xi _{\pm
}^{\theta }$ are exactly obtainable. We find that the squeezing magnitudes $%
\xi _{\pm }=\min_{\theta }\xi _{\pm }^{\theta }$ evolve in the same way for $%
{\bf J}_{+}$ and ${\bf J}_{-}$, while their squeezing directions
corresponding to the optimal angle $\theta $ are always orthogonal to each
other. This means that the criterion $\min_{\theta }\left( \xi _{+}^{\theta
}+\xi _{-}^{\theta +\pi /2}\right) /2$ for entanglement between the two
effective modes $a_{\pm 1}^{\prime }$ evolves in the same way as the
criterion $\xi _{\pm }$ for many-particle entanglement . The resultant time
evolution of $\xi _{\pm }$ is shown in Fig. 1a, together with a prediction
from the simple Hamiltonian $J_{+}^{x2}$ (or $J_{-}^{y2}$). One can see that
the results from the two approaches agree quite well, with only slight
difference in the maximum obtainable squeezing and the corresponding
evolution time. For $10^{5}$ atoms, about three orders reduction is
obtainable in $\xi _{\pm }$, which is a clear demonstration that substantial
many-particle entanglement and two-mode entanglement have been generated.
Fig. 1b shows the time evolution of the three-mode entanglement $E_{3}$. One
can see that significant three-mode entanglement can be generated with only $%
10^{3}$ atoms. 

Finally, we would like to address some imperfections and approximations used
in this scheme. First, in a practical case there would be some remaining
external magnetic field due to the special way of preparing the condensate.
It is remarkable that the linear Zeeman shift from a known homogeneous
magnetic field has no influence on our scheme since it commutes with the
spin-dependent Hamiltonian $H_{a}$ and will be eliminated if one transforms
to the interaction picture. As a result, only quadratic Zeeman shift and
linear Zeeman shift from an inhomogeneous field have influence. These
high-order effects can be easily controlled very small in the relevant time
scale. For instance, one can see from Fig. 1 that the time scale in this
scheme is roughly set by the inverse of $N\lambda _{a}^{\prime }$, which is
about $2\pi \times 10^{2}$Hz for a typical density $3\times 10^{14}$cm$^{-3}$
of sodium atoms [11]. Within this time, the quadratic and the inhomogeneous
Zeeman shifts have negligible effects if the magnetic field and its gradient
are respectively smaller than $1$mG and $1$mG/cm. Secondly, we have assumed
that the three components with $m_{f}=0,$ $\pm 1$ have the same spatial wave
function. This is a good approximation as long as there is no demixing
instability leading to phase separation of different components [2,22]. For
the two-component condensate with components $\alpha $ and $\beta $ having
equal macroscopic populations, a mean-field theory has shown that there is
no demixing instability if the inter-component collision coefficient $%
\lambda _{\alpha \beta }$ is smaller than the geometric average of the
intra-component collision coefficients $\sqrt{\lambda _{\alpha \alpha
}\lambda _{\beta \beta }}$ [22,11]. For the relevant dynamics in our scheme
with only the level $\left| m_{f}=0\right\rangle $ macroscopically
populated, Fig. 1 shows that this system is well approximated by two
independent sets of two-component condensates. For each set of two-component
condensate, we can apply the above result, and find there is no demixing
instability in this scheme if the asymmetric collision coefficient $\lambda
_{a}$ is positive, as it is for sodium atoms. This is different to the case
where both of the levels with $m_{f}=0,$ $+1$ are equally populated, which
leads to demixing instabilities [10,11]. Finally, we assumed the single-mode
approximation for the atomic motional state. This approximation is supported
by the following arguments: first, in the two-component condensate more
complete numerical simulations gives a squeezing which agrees quite well
with the calculation from the single-mode approximation [2,6,23], and the
present spin-1 system is well approximated by two independent sets of
two-component condensates; and second, a direct energy analyses in Ref. [16]
also shows the validity of the single-mode approximation for the spin-1
condensate with only the level $\left| m_{f}=0\right\rangle $ initially
populated.

In summary, we have shown how to directly generate and detect various kinds
of squeezing and entanglement in the experimentally realized spin-1
condensate. This helps for the final observation of quantum entanglement in
Bose-Einstein condensates.

This work was supported by the Austrian Science Foundation, the EQUIP, the ESF, 
the European TMR network Quantum Information, and
the Institute for Quantum Information Ges.m.b.H. L.M.D. thanks in addition
the support from the Chinese Science Foundation.

\begin{figure}[tbp]
\epsfig{file=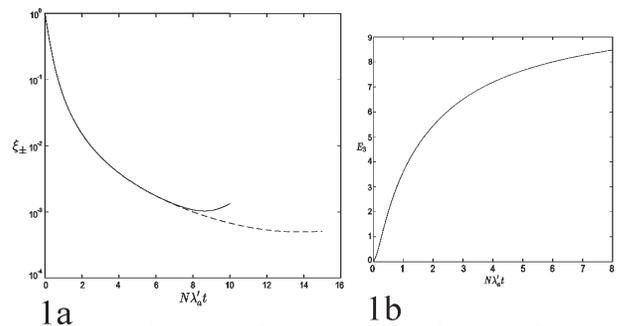,width=8cm} 
\caption{a. Time evolution of the squeezing magnitudes for $10^5$ atoms. The
solid curve shows the squeezing from an exact numerical solution to the
Hamiltonian (5), and the dashed curve is the result from the Hamiltonian $4%
\protect\lambda _{a}^{\prime }J_{+}^{x2}$. The upper dotted curve shows the
ratio of the atomic population in the level $\left| m_{f}=0\right\rangle $,
which is basically 1. b. Time evolution of the three-mode entanglement $%
E_{3} $ for $10^3$ atoms.}
\end{figure}


\begin{references}
\bibitem{21}  For instance, in the second quantization, the state $\left|
N/2,N/2\right\rangle _{+1,-1}$ is particle entangled, but not mode
entangled; and the single-particle state $\left( \left| 1,0\right\rangle
_{+1,-1}+\left| 1,0\right\rangle _{+1,-1}\right) /\sqrt{2}$ is mode
entangled, but not particle entangled.
\end{references}
\end{document}